\documentclass[]{tPHM2e}
\usepackage{bm}
\usepackage{graphicx}
\usepackage{leftidx}
\usepackage{pict2e}

\begin{document}

\title{The Sherrington-Kirkpatrick model near ${\bm T_c}$ and near
${\bm T=0}$}

 \author{A. Crisanti$^{\rm a}$$^{\ast}$%
\thanks{$^\ast$Email: andrea.crisanti@phys.uniroma1.it}%
        and C. De Dominicis$^{\rm b}$$^{+}$%
\thanks{$^{+}$Email: cirano.de-dominicis@cea.fr\vspace{6pt}}%
\\
 \vspace{6pt}
 $^{\rm a}${\em{Dipartimento di Fisica, Universit\`a di Roma 
               {\em La Sapienza} and  ISC-CNR, 
               P.le Aldo Moro 2, I-00185 Roma, Italy}};
 $^{\rm b}${\em{Institut de Physique Th\'eorique, CEA -
               Saclay - Orme des Merisiers, 91191 Gif sur Yvette, 
               France}}\\
 \vspace{6pt}
 \received{V 1.4.6 (Published) 2011/10/21 10:40:37 AC}
%
% V 1.4.3 submitted
% 
% History
%
% 1.4.4
% Added O(n^2) eq (14)
% Added sent. after eq (14) to explain ``c''
% Replaced Q(n^2, q_{ab}^5) -> O(n^2) in eq (18)
%
% 1.4.5
% small typo changes
%
% 1.4.6
% Published
%
}%

%\pacs{75.10.Nr, 64.60.De}

\maketitle

\begin{abstract}
Some recent results concerning the Sherrington-Kirkpatrick model are reported.
For $T$ near the critical temperature $T_c$, the replica free energy of the 
Sherrington-Kirkpatrick model is taken as the starting point of
an expansion in powers of $\delta Q_{ab} = (Q_{ab} - Q_{ab}^{\rm RS})$ about
the Replica Symmetric solution $Q_{ab}^{\rm RS}$.
The expansion is kept up to $4$-th order in $\delta{\bm Q}$ where a 
Parisi solution $Q_{ab} = Q(x)$ emerges, but only if one remains close enough 
to $T_c$.

For $T$ near zero we show how to separate contributions from $x\ll T\ll 1$ 
where the Hessian maintains the standard structure of Parisi Replica 
Symmetry Breaking with bands of
eigenvalues bounded below by zero modes. For $T\ll x \leq 1$ the bands
collapse and only two eigenvalues, a null one and a positive one, are found.
In this region the solution stands in 
what can be called a {\sl droplet-like} regime.
\end{abstract}

\section{Introduction}
The Sherrington-Kirkpatrick (SK) model \cite{SheKir75,SheKir78}, 
introduced in the middle of 70's as a mean-field model for spin glasses, 
has played and important role besides the study of spin glasses. 
The search for its solution in the low temperature phase, 
the spin glass (SG) phase,
has lead to the introduction and development of 
tools such as the {\sl Replica Method} and the concept of 
{\sl Replica Symmetry Breaking}, that have found applications in a variety of
other fields of the complex-system world, just to cite a few,  
neural networks, combinatorial optimization and glassy physics.

The signature of the SG state is the presence of a large number of degenerate,
locally stable states. When the system is driven into the SG phase, 
the spins almost freeze and fluctuate along local fixed directions.
Thus, similar to what happens in the low temperature phase of a ferromagnet, 
the thermal average of a single spin does not vanishes in the SG phase.
Yet, at difference with a ferromagnet, the local directions 
may change from spin to spin without any special rule, that is in a  
random way.
Therefore the SG phase is not characterized by a 
long range ordering. 
Also, since each spin can choose among different local 
directions, the SG phase is actually made by a large number of degenerate 
locally stable states. 
If we take a shot of one of such SG state, we shall discover that it 
is not distinguishable from a paramagnetic, i.e., disordered state.
Thus we can think of the SG phase made by a large set of almost frozen
paramagnetic (disordered) states. 

Direct consequence of such a scenario is that identical ideal replicas
of the system, indistinguishable in the paramagnetic phase, 
when driven into the SG phase may end up into different 
SG states and, hence, becoming distinguishable.
The symmetry among replicas is then {\sl broken}. 

The solution of the SK model which accounts for this scenario
was found at the end of the 70's, and it is
now know as the ``Parisi Solution'' \cite{Parisi79,Parisi80}.
Since then many of its properties have been studied and understood. 
Some questions, however, are still open, among them its validity for 
finite dimensional systems. 
Despite the subtleties, and mathematical ambiguities,  of the replica method,
see e.g. \cite{Dotsenko10}, it is now accepted that the Parisi Solution is
correct for the mean-field SK model. 
For finite dimensional systems it is less clear and a different scenario,
the {\sl droplet} picture, an essentially replica symmetric description,
as been put forward. The connection between the two is still not fully
understood.

In this paper we shall discuss SK model replica symmetry breaking solutions 
that are obtained by 
expanding about the replica symmetric solution. The interest for these
solutions is twofold: from one side one can investigate the set up of
the replica symmetry breaking in the SG phase. The study of these solutions
is also relevant for the definition of a field theory description of the
SG phase that could be used to investigate the crossover from 
mean-field to the finite dimensional systems.

In the last part of the paper we shall briefly discuss the solution of 
the SK model in the limit $T\ll 1$, where an (almost) replica symmetric 
description arises.

\section{The SK model and the SK-solution}
The Sherrington-Kirkpatrick (SK) model in absence of external fields 
is defined by the Hamiltonian \cite{SheKir75,SheKir78}
\begin{equation}
 {\cal H} = -\frac{1}{2}\sum_{i\not=j}^{1,N}\, J_{ij}\,\sigma_i\sigma_j
\end{equation}
where $\sigma = \pm 1$ are Ising spins and the symmetric couplings 
$J_{ij}$ are i.i.d. quenched Gaussian random variables of zero mean and 
variance 
equal to $1/N$. All pairs of spins $(i,j)$ interacts, and the scaling
of the variance ensures a well defined thermodynamic limit as $N\to\infty$.

The thermodynamic properties of the model are obtained from the free energy 
(density) $-\beta N f_J = \ln Z_J$, where $\beta = 1/T$ is the inverse 
temperature. In disordered systems $Z_J$, and hence $f_J$, is a random 
quantity.  We must therefore average the free energy $f_J$ over the disorder.
To overcome the 
difficulties of averaging a logarithm, the average over the disorder is 
computed using the so-called {\sl replica trick}. This procedure is
essentially the identity
\begin{equation}
 -\beta N f = \left[\ln Z\right]_J 
            = \lim_{n\to 0} \frac{\left[Z^n\right]_J - 1}{n},
\end{equation}
where the square brackets denote disorder average, and
$f$ the disorder averaged free energy density.
For an integer $n$, $Z^n$ may be expressed as $Z^n = \prod_{a=1}^{n} Z_a$, and
may be interpreted as the partition function of $n$ identical, 
non-interacting, {\sl replicas} of the real system. Averaging $Z^n$ for 
integer $n$ over disorder introduces an effective interaction between replicas.
Performing the average, and introducing the auxiliary symmetric replica 
overlap matrix $Q_{ab} = \frac{1}{N}\sum_i \sigma_{ia}\sigma_{ib}$,
 with $a\not=b$,  the disordered averaged replica partition function can be
written as
\cite{SheKir78}:
\begin{eqnarray}
 \left[Z^n\right]_J = \int \prod_{a<b} \sqrt{\frac{N\beta^2}{2\pi}}\, dQ_{ab}
             \, e^{N {\cal L}[{\bm Q}]}
\label{eq:lag}
\end{eqnarray}
with the effective Lagrangian (density) 
\begin{eqnarray}
{\cal L}[{\bm Q}]&=& -\frac{\beta^2}{4} \sum_{ab} Q_{ab}^2 
                       + \Omega[{\bm Q}]
                       - n\frac{\beta^2}{4}
\label{eq:l} \\
\Omega[{\bm Q}] &=&\ln {\rm Tr}_{\sigma_a} \exp\bigg(\frac{\beta^2}{2}\sum_{ab} 
             Q_{ab}\, \sigma_a\sigma_b\bigg)
\label{eq:ome}
\end{eqnarray}
The last term in (\ref{eq:l}) follows from the definition
$Q_{aa} = 1$. 
The normalization factor in (\ref{eq:lag}) gives a sub-leading 
contributions for $N\to\infty$ and is omitted in the following.

In the thermodynamic limit, $N\to\infty$, the integral over $Q_{ab}$ in 
(\ref{eq:lag}) is evaluated at the stationary point,
\begin{equation}
\label{eq:stp}
\frac{\partial}{\partial\,Q_{ab}}\,{\cal L}[{\bm Q}] = 0, \quad
a < b,
\end{equation}
that leads to the self-consistent equation for $Q_{ab}$:
\begin{equation}
\label{eq:stpeq}
  Q_{ab} 
         = \frac{{\rm Tr}_{\bm\sigma} \sigma_a\sigma_b\,
             \exp\left(\frac{\beta^2}{2}\sum_{ab} Q_{ab}\right)}
         { {\rm Tr}_{\bm\sigma}
      \exp\left(\frac{\beta^2}{2}\sum_{ab} Q_{ab}\, \sigma_a\sigma_b\right)}
= \langle\sigma_a\sigma_b\rangle, \qquad
a \not= b.
\end{equation}
The replica free energy density then reads:
\begin{equation}
\label{eq:freen}
  -n\beta\,f(n) = {\cal L}[{\bm Q}],
\end{equation}
and the average free energy $f$ is recovered as the limit 
$n\to 0$ of $f(n)$.

To solve the self-consistent stationary point equation (\ref{eq:stpeq})
an assumption on the structure of the overlap matrix $Q_{ab}$ must be done.
As the $n$ replicas of the real system are identical, one may 
reasonably assume that the solution should be symmetric under the exchange of 
any pair of replicas. Based on this, SK in their original work
assumed \cite{SheKir78}
\begin{equation}
\label{eq:qrs}
 Q_{ab} = \delta_{ab} + q\, (1-\delta_{ab})
\end{equation}
where $q$ is the overlap between any pair of different replicas.
This form of $Q_{ab}$ is what it is now known as the 
Replica Symmetric (RS) {\sl Ansatz}.
The physical meaning of $q$ is
\begin{equation}
  q = \left[ \langle \sigma_i\rangle^2 \right]_J
\end{equation}
where $\langle\cdots\rangle$ denotes here thermal average for fixed 
disorder.  
In absence of an external field the local direction of 
$\langle \sigma_i\rangle$ is random, and depends on the
disorder realization, and hence $\left[\langle\sigma_i\rangle\right]_J=0$.
Therefore a nonzero $q$ indicates local magnetic order, without a long
range ordering.

Substitution of Ansatz (\ref{eq:qrs}) into the stationary point equation 
(\ref{eq:stpeq}), or into eqs. (\ref{eq:l}), (\ref{eq:ome}) and (\ref{eq:stp}), 
leads in the $n\to 0$ limit to the self-consistent equation
\begin{equation}
\label{eq:sksol}
  q = \int_{-\infty}^{+\infty}\frac{d z}{\sqrt{2\pi q}}\
           e^{-z^2/2q}\,\tanh^2(\beta z),
\end{equation}
that, besides the paramagnetic solution $q=0$, 
admits for $T<T_c=1$ a non-trivial $q\not=0$ solution.
The order parameter $q(T)$ is a
decreasing function of temperature, that is equal to $1$ at $T=0$ and 
vanishes for $T\to T_c^{-}$.
An explicit solution of (\ref{eq:sksol}) may be obtained by expanding near the 
points $T=0$ and $T=1$. One then finds that $q$ vanishes as $T_c-T$ 
as the critical temperature $T_c$ is approached from below, 
while $1-q(T)\sim T$ as $T\to 0$.

This solution yields, however, an unphysical negative zero temperature 
entropy \cite{SheKir78}. 
The analysis of the stability of the stationary point also reveals 
that the RS stationary point becomes unstable as $n\to 0$ for all temperatures 
$T<T_c$
\cite{deAlmTho78}.

The failure of the RS Ansatz has a physical origin. 
Below the critical temperature $T_c$ 
the phase space of the SK model breaks down into a large, yet non extensive, 
number of degenerate locally stable states in which the system freezes. 
The symmetry under replica exchange is then spontaneously broken, 
and the overlap matrix 
becomes a non-trivial function of the replica indexes. Following the 
parameterization introduced by Parisi \cite{Parisi79, Parisi80}, 
the overlap matrix $Q_{ab}$ for $R$ steps of replica permutation
symmetry breaking --called RSB solution-- is divided along the diagonal 
into successive boxes of decreasing size $p_r$, with $p_0 = n$ and $p_{R+1}=1$, 
and elements given by:
\begin{equation}
Q_{ab} \equiv Q_{a\cap b=r} = Q_r, \qquad r = 0,\cdots, R+1
\end{equation}
with $1=Q_{R+1}\geq Q_R \geq\cdots \geq Q_1 \geq Q_0$.
In this notation 
$r = a\cap b$ denotes the overlap between the replicas $a$ and $b$, and 
means that $a$ and $b$ belong to the same box of size $p_r$ but to two 
{\it distinct} boxes of size $p_{r+1} < p_r$.

\begin{figure}
\begin{center}
\includegraphics[scale=1.0]{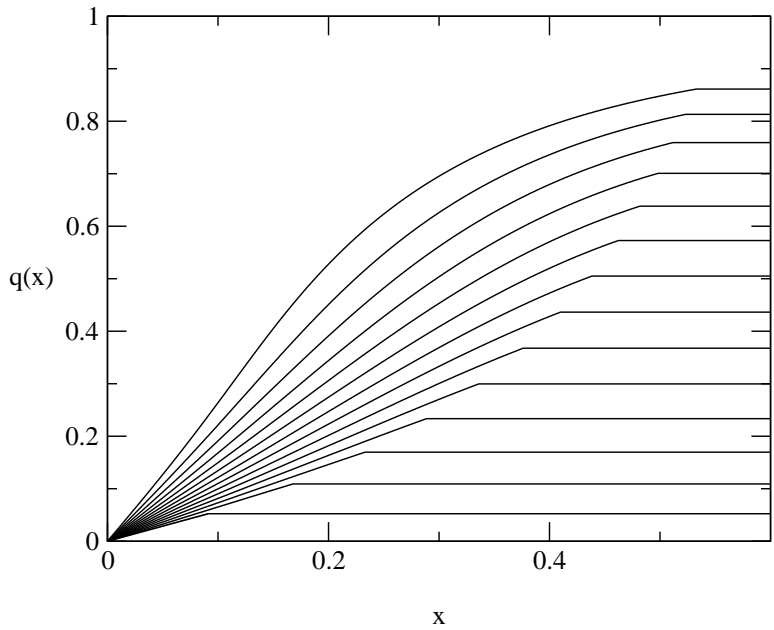}
\caption{$Q(x)$ as function of $x$ for various temperatures.
          From bottom to top $T=0.95$ to $T=0.30$ in step of
         $0.05$ [From Ref. \cite{CriRiz02}].
         }
\label{fig:qxSK}
\end{center}
\end{figure}

The solution of the stationary point equation can be obtained for any value of
$R$ \cite{Parisi79,Parisi80,CriDeDom11}.
The case $R=0$ gives trivially back the RS solution. 
It turns out that a physically acceptable solution for the SK model is 
obtained only by letting $R\to\infty$,
that is by allowing for an infinite 
number of possible spontaneous breaking of the replica permutation symmetry.
The solution is called full replica symmetry breaking (FRSB), or 
infinite-RSB solution ($\infty$-RSB) to stress the limit $R\to\infty$.
In this limit $Q_r-Q_{r-1}\to 0$, $r = 0,\ldots, R$, 
and the matrix $Q_{ab}$ is described by a continuous, non-decreasing
function $Q(x)$ parameterized by a variable $x$. The meaning of $x$ may 
depend on the parameterization used. In the Parisi scheme $x\in[0,1]$ and 
measures the probability for a pair of replicas to have an overlap not larger 
than $Q(x)$ \cite{Parisi83}.
The FRSB equations can be solved in the full low temperature
phase 
\cite{SomDup84,CriRiz02,CriRizTem03,Pankov06,OppShe05,OppSchShe07,SchOpp08}.
In Fig. \ref{fig:qxSK} we show the form of $Q(x)$ obtained from the numerical
solution of the FRSB equation [From Ref. \cite{CriRiz02}].

\section{Expansion around the RS solution}

From the analysis of the replica symmetry breaking solution at
small finite replica number $n$, Kondor \cite{Kondor83} has found 
that below, but close to, the critical temperature $T_c$ the instability 
of the RS solution appears at the finite value
$n = n_s(T) < 1$. For $n\geq n_s(T)$ the free energy $f(n)$ coincides 
with $f_{\rm SK}(n)$ from the RS solution, while for $0\leq n\leq n_s(T)$ 
it is given by 
$f_{\rm FRSB}(n)$ from the FRSB solution. 
The crossover is rather smooth since $f_{\rm SK}(n)$ and 
$f_{\rm FRSB}(n)$, along with their first two derivatives, 
coincide at $n_s(T)$.
To investigate the nature of the replica symmetry breaking in the low 
temperature phase one can then expand $Q_{ab}$ around the 
RS solution.

One then considers an overlap matrix $Q_{ab}$ of the form, 
see eq. (\ref{eq:qrs}),
\begin{equation}
\label{eq:Q}
Q_{ab} = Q_{ab}^{\rm RS} + q_{ab}
       = \delta_{ab} + q\,(1-\delta_{ab}) + q_{ab}
\end{equation}
where $q$ is given by the SK Replica Symmetric solution (\ref{eq:sksol}) and
$q_{ab}$ is the deviation from the Replica Symmetric solution. 
Inserting this form of $Q_{ab}$ into (\ref{eq:l}), and expanding the 
functional $\Omega[{\bm Q}]$ in powers of $q_{ab}$, yields:
\begin{eqnarray}
\label{eq:freenc}
-n\beta f = -n\beta f_{\rm SK} 
            &-&\frac{\beta^2}{2} q \sum_{ab}q_{ab} 
            -\frac{\beta^2}{4}\sum_{ab}q_{ab}^2 
\nonumber\\
 &+& \sum_{k\geq 1} \frac{1}{k!}\left(\frac{\beta^2}{2}\right)^k 
    \left\langle\left(\sum_{ab}q_{ab}\,\sigma_a\sigma_b\right)^k\right\rangle_c
 + O(n^2)
 \end{eqnarray}
where the subscript ``$c$'' indicates that only connected contributions
must be considered, 
i.e., only the terms that cannot be written as the product of two or more 
independent sums. All others give contributions of $O(n^2)$, or higher.
The first contribution, $f_{\rm SK}$, is the SK free energy
\begin{equation}
\label{eq:freenSK}
-\beta f_{\rm SK} = \frac{\beta^2}{4}q^2 - \frac{\beta^2}{2}q 
           + \overline{\ln\cosh(\beta z)} + \ln 2
             + O(n)
\end{equation}
where the overbar denotes the average over the Gaussian variable $z$:
\begin{equation}
\overline{g(z)} = \int_{-\infty}^{+\infty}\frac{d z}{\sqrt{2\pi q}}\
           e^{-z^2/2q}\,g(z).
\end{equation}
Whit this notation the RS solution (\ref{eq:sksol}) takes the compact
form
\begin{equation}
\label{eq:qSK}
  q = \overline{\theta^2}, \qquad \theta \equiv \tanh(\beta z).
\end{equation}

Expanding the cumulants in eq. (\ref{eq:freenc}) up to order $O(q_{ab}^4)$ 
included, the lowest term needed to break replica symmetry, one obtains 
\cite{CriDeDom10}
\begin{eqnarray}
\label{eq:freen4c}
-n\beta f &=& -n\beta f_{\rm SK} 
  +\frac{1}{4T^4}
            \left[M\,\sum_{abc}q_{ac}q_{cb} + N\, \sum_{ab}q_{ab}^2\right]
\nonumber\\
  &\phantom{=}&
   +\frac{1}{6(2T^2)^3}
            \left[
    P\,\sum_{abcd}q_{ac}q_{cd}q_{db} 
  + Q\,\sum_{abcd} q_{ad}q_{bd}q_{cd}
  + R\,\sum_{abc} q_{ac}^2q_{cb} 
    \right.
\nonumber\\
  &\phantom{=}&\hspace{5.5cm}
    \left.
  + J\,\sum_{ab}q_{ab}^3 
  + K\, \sum_{abc} q_{ac}q_{cb}q_{ba}
            \right]
\nonumber\\
  &\phantom{=}&
  +\frac{1}{24(2T^2)^4}
            \left[
   - A\,\sum_{abcde}q_{ae}q_{be}q_{ce}q_{de}
   + B\,\sum_{abcd} q_{ac}q_{cd}^2q_{db}
   \right.
\\
  &\phantom{=}&\hspace{1.5cm}   
   - B\,\sum_{abcde}q_{ac}q_{dc}q_{ce}q_{eb}
   + C\,\sum_{abc} q_{ac}q_{cb}^2q_{ba}
   - C\,\sum_{abcd}q_{ac}q_{ad}q_{dc}q_{cb}
\nonumber\\
  &\phantom{=}&\hspace{1.5cm}   
   +4D\,\sum_{abc}q_{ac}^3q_{cb}
   -3D\,\sum_{abcd}q_{ab}^2q_{bc}q_{bd}
   +E\,\sum_{abcde}q_{ab}q_{bc}q_{cd}q_{de}
\nonumber\\
  &\phantom{=}&
   \left.
   -2E\,\sum_{abcd}q_{ab}q_{bc}q_{cd}^2
   +F\,\sum_{abcd}q_{ab}q_{bc}q_{cd}q_{da}
   + G\,\sum_{ab}q_{ab}^4 
   - H\,\sum_{abc}q_{ac}^2q_{cb}^2
   \right]
\nonumber\\
  &\phantom{=}&     
+ O(n^2)
\nonumber
\end{eqnarray}
where
\begin{equation}
M = 2\,\overline{\theta^2(1-\theta^2)}, \quad
N = \overline{(1-\theta^2)^2} - T^2.
\end{equation}
\begin{equation}
    P =  24\,\overline{\theta^2(1-\theta^2)^2}  ,\quad
    Q = -16\,\overline{\theta^4(1-\theta^2)}    ,\quad
    R = -48\,\overline{\theta^2(1-\theta^2)^2}  ,\quad
\end{equation}
\begin{equation}
    J =  16\,\overline{\theta^2(1-\theta^2)^2}  ,\quad
    K =   8\,\overline{(1-\theta^2)^3}
\end{equation}
\begin{equation}
A =  32\,\overline{\theta^4(1-3\theta^2)(1-\theta^2)}, \quad
B = 384\,\overline{\theta^4(1-\theta^2)^2}, \quad
C = 384\,\overline{\theta^2(1-\theta^2)^3},
\end{equation}
\begin{equation}
D =  64\,\overline{\theta^2(1-3\theta^2)(1-\theta^2)^2}, \quad
E = 192\,\overline{\theta^2(1-\theta^2)^2}, \quad
F =  48\,\overline{(1-\theta^2)^4},
\end{equation}
\begin{equation}
G =  32\,\overline{(1-3\theta^2)^2(1-\theta^2)^2}, \quad
H =  96\,\overline{(1-3\theta^2)(1-\theta^2)^3}.
\end{equation}

\begin{figure}
\begin{center}
\includegraphics[scale=1.0]{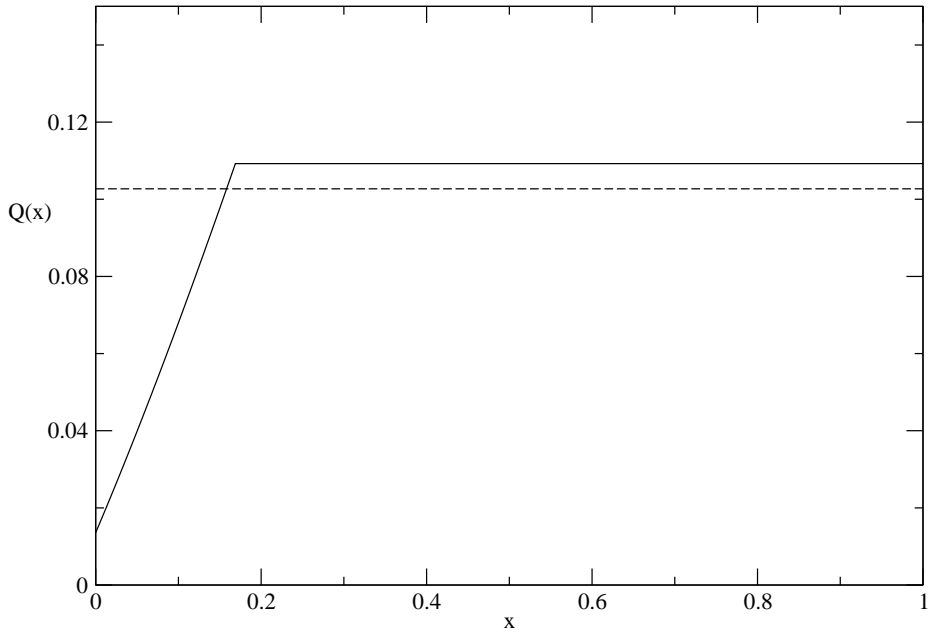}
\caption{$Q(x)$ versus $x$ at temperature $T=0.9$. The horizontal dashed line
         shows the SK solution $Q(x) = q$. For this 
         temperature we have $x_c = 0.168846\ldots$, $Q(x_c) = 0.109238\ldots$ 
         $Q(0) = 0.013570\ldots$ and $q = 0.102701\ldots$. 
         }
\label{fig:q09}
\end{center}
\end{figure}

The equation for $q_{ab}$ follows from the stationarity condition 
$(\partial/\partial q_{ab}) f = 0$ applied to the replica free energy 
functional
(\ref{eq:freen4c}).  In the limit $R\to\infty$ this becomes an 
integro-differential equation for $q(x)$, whose solution is 
\begin{equation}
\label{eq:qx}
q(x) = \Gamma \frac{x-s}{\sqrt{(x-s)^2 + \Delta}} 
      -a - bS_1, \qquad
0\leq x\leq x_c,
\end{equation}
and $q(x) = q(x_c)$ for $x_c \leq x <1$.
The parameters $a$, $b$, $s$ and $\Delta$ are functions of the coefficients
of the expansion (\ref{eq:freen4c}), and hence of $T$, 
while $\Gamma$, $S_1$ and $x_c$ must be determined selfconsistently. 
For details we refer to Ref. \cite{CriDeDom10}.
It turns out that below the temperature $T=0.549\ldots$ no physical solution 
with $q(x)\not= 0$ is exists and only the SK solution $Q(x) = q$ survives.
 Figure \ref{fig:q09} shows the solution $Q(x) = q + q(x)$ for $T=0.9$.

Near the critical temperature $T_c=1$ the solution can be expanded in powers of
$\tau = T_c-T \ll 1$. The first terms of the expansion of $q(x)$ 
for $0 \leq x\leq x_c$ read
\begin{eqnarray}
 q(x) &=& \frac{x}{\sqrt{4+6x^2}}
        + \left[1 - \frac{4\sqrt{2}}{(2+3x^2)^{3/2}} + 
                \frac{3x(2+x^2)}{\sqrt{2}(2+3x^2)^{3/2}}
          \right]\,\tau
\nonumber\\
&\phantom{=}& +
\frac{1}{3(2+3x^3)^{5/2}}\Bigl[
   64\sqrt{2}     
  -236\sqrt{2}x + 312\sqrt{2}x^2 - 138\sqrt{2}x^3 - 81\sqrt{2}x^5 
\nonumber\\
&\phantom{=}&\phantom{=======}
     - 56 \sqrt{2+3x^2}  - 168 x^2\sqrt{2+3x^2} - 126x^4\sqrt{2+3x^2}      
         \Bigr]\, \tau^2
\nonumber\\
&\phantom{=}&
+ O(\tau^3),
\end{eqnarray}
while for the breaking point $x_c$ one has 
\begin{equation}
  x_c = 2\tau - 4\tau^2 + \frac{40}{3}\tau^3 - \frac{665}{9}\tau^4
        + \frac{68567}{135}\tau^5 + O(\tau^6).
\end{equation}
From these expressions, and the expansion of the RS solution $q$ close 
to $T_c$, follows
\begin{equation}
\label{eq:qx0}
Q(0) = q + q(0) =  \frac{56}{3}\tau^3  
                 - \frac{220}{3}\tau^4 
                 + \frac{3968}{9}\tau^5
                 + O(\tau^6) 
\end{equation}
\begin{equation}
Q(x_c) = q + q(x_c) =   \tau
                      + \tau^2 
                      - \tau^3 
                      + \frac{5}{2}\tau^4
                      - \frac{413}{90}\tau^5
                      + O(\tau^6).
\end{equation}
A peculiarity of this FRSB solution is that $Q(x)$ does not vanishes at $x=0$,
as Fig. \ref{fig:q09} clearly shows.
$Q(x=0)$ only vanishes at $T_c$, and grows as $(T_c - T)^3$,
see eq. (\ref{eq:qx0}), below it.
The reason for such a behaviour is that the FRSB ``opens'' 
around the RS solution $Q^{{\rm RS}}(x) = q$.
As the temperature decreases the RS solution $q$ grows from zero 
and drags $Q(x=0)$ to finite values. At $T=0.618\ldots$ 
the value of $Q(x=0)$ eventually overcomes that of $q$.

In the absence of external fields that break the up/down symmetry 
$Q(x=0)$ must vanish. The finite value of $Q(x=0)$ then indicates that more
terms in the expansion (\ref{eq:freenc}) must be retained in order to 
balance the drag from $q$.
Alternatively a null value of $Q(x=0)$ can be recovered by taking $q=0$, 
which  eliminates the drag from the beginning.
Physically this means by performing 
the expansion around the paramagnetic solution.
In this case, however, the FRSB solution exists in a narrower region,
$0.9148\ldots\leq T\leq 1$, close to the critical temperature.

We note that the choice $q=0$ was used to derive the so called 
{\sl Truncated Model} \cite{Parisi79,BraMoo79} 
largely used to study the low temperature phase of the SK model.
This model is based on an expansion that
retains only the main mathematical structure of the expansion of the 
replicated free energy in powers of $Q_{ab}$ near $T_c$, where $|Q_{ab}|\ll 1$,
but, similar to the Landau Lagrangian, with arbitrary coefficients.
The mapping between the expansion (\ref{eq:freen4c}) with $q=0$ 
and the Truncated Model is 
$F=H=0$, $N=2T^3 (1-\tau)$, $K=8T^5w$ and $G=32T^7u$  with arbitrary $u$ and 
$w$.
In this case, if $w$ and $u$ are temperature independent, the FRSB solution 
exists only down to $T_{\rm trm} = 1 - w^2/4u$.

\section{Solution near $T=0$}
In the previous Section we have seen that the RS solution can be taken as 
a starting point for a development of the FRSB solution for $T_c-T\ll 1$.
As the temperature becomes significantly lower than $T_c$ the expansion looses 
physical meaning and more and more terms are needed into the expansion to 
extend its validity. Due to the particular structure of the 
expansion, in powers of $\beta$, it is likely that an infinite subset of 
terms should be considered to tackle with very low temperatures.

\begin{figure}
\begin{center}
\includegraphics[scale=1.0]{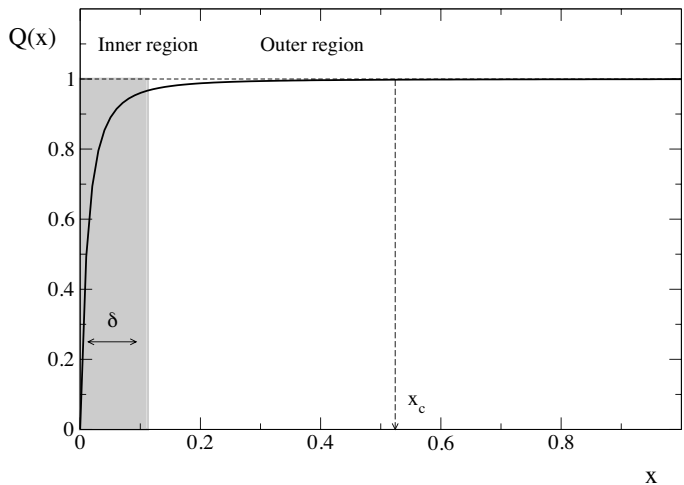}
\caption{Shape of $Q(x)$ as function of $x$ for $T\ll 1$. 
         The shaded ares shows the extent of the layer of
         thickness $\delta\sim T$ where the larger variation
         of $Q(x)$ is concentrated.
         }
\label{fig:qx}
\end{center}
\end{figure}

Surprisingly a Replica Symmetric description appears at $T\to 0$.
Indeed as the temperature is lowered towards $T=0$ 
the probability that $Q_{ab}\leq q_c(T) = O(1)$ remains finite, {\sl viz.}
$x_c\simeq 0.524\ldots$ as $T\to 0$ \cite{Som85,CriRiz02} . 
At the same time the probability of finding overlaps $Q_{ab}$ 
{\sl significantly smaller} than
$q_c(T) = O(1)$ vanishes with $T$. 
Then, since $Q(x=0)=0$, for $T\ll 1$ the order parameter function $Q(x)$
undergoes an abrupt and rapid change of $O(1)$ in a tiny boundary layer of 
thickness $\delta\sim T$ close to $x=0$,
while it is slowly varying for $\delta\ll x\leq x_c$, 
as depicted in Fig. \ref{fig:qx}.
In the limit $T\to 0$ the thickness $\delta\sim T\to 0$, and the order parameter
function becomes discontinuous at $x=0$.

Uniform approximate solutions valid for $T\ll 1$ can be constructed by studying
the problem separately inside ({\sl inner region}) and outside 
({\sl outer region}) the boundary layer.

The {\sl inner solution} for $Q(x)$ was first computed by 
Sommers and Dupont \cite{SomDup84}, 
and recently extensively studied by Oppermann, Sherrington 
and Schmidt \cite{OppShe05,SchOpp08,OppSchShe07}.
It turns out that as $T\to 0$ the inner solution $Q(a)$ remains  
a smooth function the inner variable $a=x/T$, needed to blow up the 
inner region, varying between $0$, for $a=0$, and $q_c(T)\sim 1$,
for $a\gg 1$.
The {\sl outer solution} was studied by Pankov \cite{Pankov06}, 
who found that in the outer region one has
\begin{equation}
\label{eq:pank}
  Q(x) \sim 1 - c(\beta x)^{-2}, \qquad
T\ll x \ll x_c\ \mbox{\rm and}\ T\to 0
\end{equation}
where $c=0.4108\ldots$ and $x_c\simeq x_c(T=0)= 0.524\ldots$.

This behaviour of $Q(x)$  has strong consequences on other relevant 
quantities and, also, on the structure of the eigenvalue spectrum of
the Hessian of the fluctuations governing the stability of the stationary
point.
For what concerns the stability one finds \cite{CriDeDom10a,CriDeDom11} 
that in the region
$x\leq \delta \sim T$, where $Q(x)$ varies rapidly from $0$ up to
$q_c\simeq 1$, the spectrum of the Hessian of the fluctuations maintains 
the complex structure of the FRSB solution found close to 
$T_c$ \cite{DeDomKon83}. 
That is a Replicon 
band whose lowest eigenvalues are zero modes, and a Longitudinal-Anomalous
band of positive masses.

In the region $T\ll x\leq x_c$, the eigenvalue spectrum has a completely 
different aspect. The bands observed in the FRSB regime collapse and only 
two distinct eigenvalues are found: a null one and a positive one 
\cite{CriDeDom10a}.
This ensures that the FRSB solution remains stable down to zero temperature.
We note that zero modes arise from the Replicon geometry, with Ward-Takahashi 
identities protecting them \cite{DeDomTemKon98},
and arises also from the Longitudinal-Anomalous 
geometry, without protection of the Ward-Takahashi identities.
It is worth to observe that the stability analysis of the RS solution also 
leads to two eigenvalues, one of which is zero to the lowest order in $T_c-T$
(and negative to higher order) and a positive one \cite{deAlmTho78}.

From the expression (\ref{eq:pank}) it follows that the variation of $Q(x)$ 
in the outer region, $T\ll x\leq x_c$, is 
$[Q(x_c) - Q(x)] / Q(x) \sim c (T/x)^2$, rather weak for $T\to 0$.
Thus in this region, that covers the overwhelming part of the interval 
$0 < x < 1$ for $T\to 0$,  we have a marginally stable (almost) Replica
Symmetric solution, that become a genuine Replica Symmetric solution for 
$T\to 0$, with self-averaging trivially restored.

\section{Discussion}
In this paper we have reported some recent results on the solution of the
SK model.  In particular we have considered two different issues.
In the first part we discussed the analysis of the Replica Symmetry Breaking 
done through an expansion around the Replica Symmetric solution.
In the second part we examined the properties of the solution in the limit 
of vanishing temperature.

The expansion of the replica free energy functional around the RS solution,
truncated to the  fourth order in $q_{ab} = Q_{ab} - Q_{ab}^{\rm RS}$,
leads to a FRSB solution with a continuous order parameter $Q(x)$
below the critical temperature  $T_c=1$. 
The solution, however, exists only in the range 
$0.549\ldots\le T\le T_c$. Moreover the value of $Q(x=0)$ is finite 
and vanishes only for $T\to T_c^{-}$ with the third power of the temperature
difference.
A null $Q(x=0)$ can be recovered by expanding about the (replica symmetric)
paramagnetic solution $q=0$. In this case, however, the FRSB solution
exists only close to $T_c = 1$, in the narrower interval 
$0.915\ldots\le T\le T_c=1$.

For what concerns the stability we believe that the solution form 
the expansion around the paramagnetic solution $q=0$ has the same 
stability properties in the most dangerous sector, i.e., in the 
Replicon subspace by virtue of the Ward-Takahashi identities
\cite{DeDomTemKon98}, of the Truncated Model.
We believe that this feature remains true for the expansion around the 
SK solution as well \cite{CriLeuParRiz04}. Thus the FRSB, where it does exist,
is marginally stable with null Replicon eigenvalues.

Finite RSB solutions, or even RS solution $q_{ab} = q\, (1-\delta_{ab})$, 
may also exists. These may exists for all temperatures below $T_c$, or only 
in a limited range of temperature. In either case all these solutions are 
unstable, with negative Replicon eigenvalues.

The main limitation of the expansions discussed here is the limited range 
of temperatures where the FRSB solution exists. To extend the range 
one should retain more terms in the expansion. This will also cure the
finite value of $Q(x=0)$.
Due to the particular structure of the expansion, in powers of $\beta$, it
is likely that an infinite subset of terms should be considered to extend
the validity of the expansion to very low temperatures.
If one is interested into these temperatures different approaches, e.g., 
that proposed in Ref. \cite{CriDeDomSar09} based 
upon an expansion around a spherical 
approximation that leads instead to an expansion in $T$,
may be more suitable.

The striking property of the solution of the SK model for $T \ll 1$ 
is the presence of two well distinct regions where the order parameter function 
$Q(x)$ behaves differently. 
In the first region, close to $x=0$, 
$Q(x)$ varies rapidly from $Q(x=0)=0$ to $Q(x)\sim q_c=O(1)$ for
$x\sim T$. 
In this region, that contains (almost) the whole variation of $Q(x)$,
the solution maintains the structure of the FRSB solution
found for higher temperature and close to $T_c$, including the
complex Hessian spectrum, even in the limit $T\to 0$.
For this reason this region was called the {\sl RSB-like} regime.

In the second region, for $T\ll x\leq x_c\simeq = 0.575\ldots$, 
$Q(x)$ is a very slow varying 
function of $x$, the variation being indeed of the order 
$[Q(x_c)-Q(x)]/Q(x) = {\cal O}\left((T/x)^2\right)$.
Here the bands observed in Hessian spectrum for the RSB regime disappear,
and only two distinct eigenvalues are found: a null one and a positive
one, ensuring the stability of the FRSB solution down to $T\to 0$.
In this region, that covers the overwhelming part of the interval  $0< x < 1$
for $T\ll 1$, the solution strongly resembles a stable RS  solution
typical of a droplet description. For this reason this region was called
{\sl droplet-like} regime.

In the limit $T\to 0$ the domain of the RSB-like regime shrinks to zero, 
and only the droplet-like part of the solution remains. It is then tempting
to interpret this as a zero temperature transition/crossover from a 
FRSB solution 
to a droplet scenario. We stress, however, that while 
these results strongly suggest a transition or crossover 
between RSB and droplet descriptions
in spin glasses, to have a better understanding of the behavior
of finite dimensional systems loop corrections to the mean-field 
propagators must be considered \cite{BraMoo86}.
Work in this direction is in progress.

\end{document}